\documentstyle[12pt]{article}
\begin {document}
\begin{flushright}
OITS-567\\
February 1995\\
\end{flushright}

\begin{center}
{\large {\bf CLUSTER PRODUCTION WITH \\ \vskip .35 cm COALESCENCE AND
BREAKUP}}
\vskip .75cm
\centerline{\bf Rudolph C. Hwa$^1$\ \  {\rm and}\ \  Jicai\ Pan$^2$}
\bigskip
 \centerline{{$^1$}Institute of Theoretical Science and Department of
Physics}
\centerline{University of Oregon, Eugene, Oregon 97403}
\medskip
\centerline{{$^2$} Department of Physics, McGill University}
\centerline{ Montreal, Quebec, Canada  H3A 2T8}

\vskip1cm
{\bf Abstract}
\vskip.2cm
\end{center}

\begin{quote}
\baselineskip=12pt
\quad \quad {\small The problem of hadronic cluster production in
quark-hadron
phase transition in heavy-ion collisions is studied by cellular automata.
Previous
result on the scaling behavior is extended to include variation in the
drift speed.  It is
also shown that coalescence is more important than growth in generating
scaling.  A
new set of rules is adopted to free the clusters from being rigid.  It is
found that the
scaling exponent is independent of not only the shapes of the clusters, but
also
the probability of breakup of the clusters.  The universality of the
scaling behavior
is now extended to a wide range of physical properties  characterizing the
geometry and
dynamics of the phase transition process.}
\end{quote}

\section{Introduction}

The problem of cluster production in the mixed region in heavy-ion
collisions has
been investigated during the past year, yielding the intriguing result that
the
distribution of cluster sizes exhibits universal scaling behavior
\cite{hlp,hp}.
During the mixed phase the quark-hadron system evolves by itself to
self-organized criticality (SOC)
\cite{btw}.  The method of investigation is by use of cellular automata.
Simple rules
are adopted to describe the dynamical process of formation, growth,
coalescence,
breakup and emission of the hadronic clusters in the quark-gluon plasma
undergoing
quark-hadron phase transition.  The breakup process is considered in the
1-dimensional (1D) problem \cite{hlp}, because it is easy to implement, but
not in 2D
\cite{hp} since there are too many possibilities that are difficult to
cover by a set of
simple rules.  In this paper we modify the cellular automaton so that not
only is
breakup made feasible simply, but also is the deformation of the clusters
allowed as
they evolve.

Despite the drastic simplification of the complicated nonequilibrium
dynamical process
by the adoption of a set of simple rules, an essential aspect of SOC that
renders the
result interesting physically is that the scaling behavior is independent
of the
parameters in the rules.  Moreover, the result should be independent of the
details of
the rules themselves so that the power law obtained is a general
consequence of the
physical system and not of the particular ways in which the dynamics is
transcribed
into cellular automata.  For that reason a change of rules, as considered
in this paper,
is an important part of the overall program to determine the generic
properties of SOC
that the quark-gluon system possesses.

There are three areas of investigations in this paper.  First, in the
framework of the
original cellular automaton \cite{hp}, we study the dependences on the
drift speed and
the relative importance of growth versus coalescence.  Second, we modify
the rules to
allow for deformation of the cluster shapes.  Third, using the new cellular
automaton
we consider the breakup of clusters.

\section{Drift, Growth and Coalescence}

We begin with a brief summary of the modelling of the problem and the
original rules
of the cellular automaton used in Ref. \cite{hp}.  Starting with the
conventional picture
of heavy-ion collision, there is a cylinder of quark-matter, expanding
mostly in the
longitudinal direction, and at a lower rate transversely.  Consider a cross
section of the
cylinder at midrapidity some time after the collision so that in the ideal
case there is a
circular core region containing quarks and gluons.  Hereafter, for brevity
we shall just
call them quarks with gluons implied, and refer to the core as the $Q$
region.
Surrounding the $Q$  region is the $M$ region, an annular ring in which
quarks and hadrons
coexist in the mixed phase.  Due to radial expansion the hadrons move
outward and
leave the $M$ region, while the quarks remain the $M$ region because of
confinement.  The
$Q$ region reduces in size as quarks are fed into the $M$ region; the $M$
region also
shrinks in time as hadronic clusters are emitted.  A wedge of the $M$
region is mapped
to a $L \times  L$ lattice initially, where the left boundary represents
the boundary
between the $Q$  and $M$ regions, and is fixed.  The right boundary moves
semilocally to
the left depending on where the clusters are emitted.  Periodic boundary
condition is
imposed on the upper and lower sides.

There are three parameters in the problem:  $L$, the initial lattice size
in units of some
hadronic scale; $S_0$, the nucleation size in units of the same scale; and
$p$, the
nucleation probability in each time step.  The rules of the cellular
automaton are in
essence the following.

\begin{description}
\item[a] \quad {\it Nucleation.}  An unoccupied site represents the quark
phase, while an
occupied site represents the hadron phase.  At each time step an unoccupied
site has
probability
$p$ of becoming an occupied site, which, if it is an isolate side,
constitutes a single-site
nucleation with $S_0 = 1$.
\item[b] \quad {\it Growth.}  If the newly nucleated site is a near
neighbor of an existing
occupied site, it is regarded as a growth process and the sites are bonded
to form a cluster.
Further growth of the cluster can take place in the same way in later
steps.
\item[c]\quad  {\it Drift.}  A cluster moves to the right one step plus a
random step
to any one of its four immediate neighbors.
\item[d] \quad {\it Coalescence.}  When two clusters collide, i.e., when
their constituent
sites overlap, they are bonded to form a  larger cluster with one of the
overlapping sites
moved to the nearest unoccupied site.
\item[e] \quad   {\it Emission.}  Upon reaching the right boundary a
cluster is removed
from the $M$ region, whose new boundary is reset at one layer to the left
for the same
number of vertical sites as the number of sites $S$ contained in the
cluster emitted.  When
the right boundary reaches the left boundary, the transition process is
terminated.
\end{description}

According to the above rules the distribution of the sizes of the emitted
clusters is
found to satisfy the scaling law \cite{hp}
\begin{eqnarray}
P(S) \propto S^{-\gamma} \quad ,
\label{1}
  	\end{eqnarray}
where
   \begin{eqnarray}
\gamma= 1.86 \pm 0.18  \quad .
\label{2}
  	\end{eqnarray}
Furthermore, it is found that $\gamma$ is independent of $L$ and $S_0$
(provided that
$L$ is large enough, e.g., 16 and 32), and depends very little on $p$
(provided $p$ is in
the range $0.05 \leq p \leq 0.5$).  In that sense we regard the result as
exhibiting
universal scaling behavior.

There are two questions that can be clarified without changing the rules.
One is the
dependence of the above result on the average drift velocity. The other is
the relative
importance of growth versus coalescence in their contributions to the
formation of
large clusters.  For the former we double the average drift step to 2,
keeping $S_0 = 1$
and $L = 16$, and obtain the result as shown in Fig. 1.  We see that $P(S)$
fails to
develop scaling behavior for small $p$, but for $p > 0.1$ there is scaling
with
essentially the same slope as before.  Evidently, as the clusters move more
rapidly out
of the $M$ region, they have less time to grow and coalesce with the
consequence that
for $p \leq 0.05$ the formation of large clusters is suppressed.  When we
double $L$ to 32,
then scaling is restored for $p \geq 0.05$ as indicated in Fig. 2, thus
verifying the
importance of having enough steps in the $M$ region for the large clusters
to form.
Comparing Fig. 2 to the result of the ``standard'' case $S_0 = 1$, $L =
16$, drift step = 1,
considered in \cite{hp}, and reproduced in Fig. 3, one finds that $P(S)$
are nearly
identical for each value of $p$.  Note that this is not merely a
consequence of scaling up
the lattice spacing.  The random-walk step size has not been changed.  Thus
Fig. 2
corresponds to the standard case with half the step size in random walk.
That reveals
another feature of the universality, i.e., the independence on the
temperature (if
thermal description is appropriate) at which the transition takes place.
But scaling
follows only when the $M$ region is large enough (relative to the drift
step) which is a
conclusion already reached in Ref. \cite{hp}, when $L$ is reduced to 8.

Concerning the relative importance of growth and coalescence, we consider
in the
standard case when only growth is allowed (Fig. 4) and then when only
coalescence is
allowed (Fig. 5).  It is clear from these two figures that coalescence is
the principal
cause of the scaling behavior.  Without coalescence there are no large
clusters no matter
how large $p$ is.

\section{Modified Rules}
According to the rules thus far considered, the clusters develop irregular
shapes
because the bondings are rigid. Since dendritic structure has more
unoccupied
neighboring sites, it favors growth and coalescence.  One may question
whether
that contributes to a bias in favor of developing larger clusters.  If the
surface
tension is negative, the clusters should have irregular surfaces.  But if
it is
positive and large, then they have circular shape in 2D.  It is difficult
to require
specific shapes for the clusters on a lattice by adding new rules to the
cellular
automaton used above.  We propose to modify the relevant part of the rules
fundamentally.

Our new approach is to stack a cluster vertically on only one site and to
assign an
effective radius $R$ to that site with
   \begin{eqnarray}
R = \alpha \sqrt{S / \pi}
\label{3}
  	\end{eqnarray}
where $\alpha$ is a parameter that quantifies the irregularity of the
cluster
shape.  If $\alpha = 1$, the shape is circular; if $\alpha > 1$, then it
can be dendritic
or even porous, depending on how large $\alpha$ is.  The stack is always
placed at
a lattice site, whether or not the effective circle corresponding to
(\ref{3}) covers $S$
discrete sites.  In drift and random walk the stack moves in discrete steps
as before.
Growth and coalescence can now be treated on the same footing.  Let the
distance between
two clusters, $S_1$ and
$S_2$, located at
$\left( x_1, y_1
\right)$ and
$\left( x_2, y_2\right)$, respectively, be
   \begin{eqnarray}
d_{12} = \left[ \left( x_1 - x_2 \right)^2 + \left(y_1 - y_2
\right)^2\right]^{1/2} \quad .
\label{4}
  	\end{eqnarray}
If the two sites satisfy the condition
  \begin{eqnarray}
d_{12} \leq  R_1 + R_2  \quad ,
\label{5}
  	\end{eqnarray}
then coalescence occurs.  If one of the sites is a newly occupied position
with size
$S_0$, then it is a growth [when (\ref{5}) is satisfied with the
corresponding $R$
being given by (\ref{3}) with $S = S_0$].  In either case we put the
combined cluster of
size $S_1 + S_2$ at the site closest to the center-of-mass of the two
original sites.
If more than two clusters coalesce simultaneously, we stack them all at the
new
site closest to the overall cm position.  Since we impose periodic boundary
condition on the upper and lower edges of the lattice, there are always two
vertical distances for every pair of $y_1$ and $y_2$: $\left|  y_1 -
y_2\right|$ and
$L - \left|  y_1 - y_2\right|$.  In (\ref{4}) we use the lesser of these
two for
$\left( y_1 - y_2 \right)$.

The result obtained with this new cellular automaton is shown in Fig. 6 for
the
standard case:  $L = 16$, $S_0 = 1$ and $p = 0.2$.  For $\alpha = 1.0$ the
scaling
exponent is nearly identical to the result of the earlier cellular
automaton, shown in
Fig. 3.  Thus our clusters produced previously \cite{hp} are not very
different
from circular clusters.  More significantly, the result does not depend
sensitively
on the specific rules adopted.  This is of great importance if the scaling
behavior
that emerges is to be regarded as a general consequence of the physical
system,
not of the particular cellular automaton used.  Fig. 6 also shows the two
other cases
with $\alpha = 1.2$ and $1.4$.  There is a slight increase of the slopes,
but very
little.  Note that with $\alpha = 1.4$ the effective area of a cluster is
double that
for $\alpha = 1.0$.  This enlargement causes crowding of the $M$ region and
increases the emission rate, thus reducing the probability of incubating
large
clusters.  Since the increase of the slope is so small, we should regard
the result
as essentially indicating approximate independence on $\alpha$, or the
shapes of
the clusters.

\section{Breakup}

In the original cellular automaton of Sec. 2, it is difficult to introduce
breakup,
since no simple rule can be added to take care of all possible ways a big
cluster
of irregular shape fragments.  With our new cellular automaton of Sec. 3,
the
process is easy to implement.

Our procedure is to regard breakup as a consequence of spontaneous fission
$S
\rightarrow S_1 + S_2$.  Breakup in a collision is a combination of
coalescence and
fission: $S_1 + S_2 \rightarrow S_3 \rightarrow S_4 + S_5$.  If one of
$S_4$ and $S_5$ further fissions in the next time step then the overall
reaction
may be regarded as a $2 \rightarrow 3$ cluster production process.  In each
time
step we have all four processes: nucleate, walk, coalesce, and break up, in
that
order.

Since a cluster is now a stack at one site with an effective radius of
extension,
breakup is a simple partition of the stack $S$ into two stacks $S_1$ and
$S_2$.
The only essential input is the probability of occurrence for various
channels of
partitioning.  Our guidance is the minimization of the
circumference-to-area
ratio.  Thus we parameterize the breakup probability by
   \begin{eqnarray}
B \left( S, S_1, S_2 \right) = \beta \left( \sqrt{S_1/S} +
\sqrt{S_2/S}\right)^{-b}
\label{6}
  	\end{eqnarray}
with $S_1 + S_2 = S$; $\beta$ and $b$ are both positive.  The positions of
the
$S_1$ and $S_2$ clusters are specified as follows.  If the original site of
$S$ is at
$\left( x, y \right)$, then $S_1$ and $S_2$ are placed at $\left( x, y +
y_1 \right)$
and $\left( x, y - y_2 \right)$, respectively, where $y_{1,2}$ are the
largest
integers less than $\alpha \sqrt{S_{2, 1}/ \pi } + 1$.  According to this
rule the
effective areas of $S_1$ and $S_2$ will never overlap.  If either one of
those
areas overlaps with the effective area of any existing cluster, then the
breakup is
forbidden from taking place, since overcrowding will lead to immediate
coalescence and further breakup, {\it ad infinitum}.

In our simulation we have set $b = 1$ as a representative value, and varied
$\beta$ from $0$ to $1.0$ with the other parameters set at the standard
values:
$L = 16$, $S_0 = 1$, $p = 0.01$ to $0.2$, $\alpha = 1$.  In Fig. 7 we show
the result for
$p = 0.01$.  Recall from Fig. 3 for no breakup that scaling cannot be
achieved at
the low  nucleation probability of $p = 0.01$.  Now with breakup it is even
more
difficult as indicated by the curves for $\beta = 0.5$ and $1.0$ in Fig. 7.
 Clearly,
large clusters cannot easily be formed when there are few existing clusters
to
coalesce, and even when a  large cluster comes into being, it breaks up
when
$\beta$ is large.  The situation changes as $p$ is increased, as can be
seen in Fig.
8 and 9.  When $p = 0.05$, more growth and coaslescence can occur to give
scaling $P(S)$, if $\beta = 0$.  But as $\beta$ increases, breakup still
has a small effect
in disrupting the formation of large clusters, as is evident in Fig. 8.
However, when
$p = 0.2$, Fig. 9 shows that breakup has no more effect.  There is
sufficiently
many clusters in the $M$ region that breakup and coalescence occur at equal
rate to result in negligible difference for any $\beta$.

 From these results we may conclude that when scaling is fully developed for
$0.05 < p < 0.5$ the cluster distribution is essentially independent of
whether the
clusters can break up.  Thus the universality of the scaling behavior is
broadened
in one additional aspect of the dynamics of cluster formation.

\section{Conclusion}

The complicated process of quark-hadron phase transition cannot be treated
reliably by any analytical method, especially when thermal equilibrium is
not
assumed.  The use of cellular automata, on the other hand, could also be
subject
to the criticism that the process has been approximated by a  scheme
of unknown reliability.  However, complications have not been ignored; they
have been taken into account by simple rules.  It is when the result
exhibits
independence on the details of the rules that one develops confidence in
the
meaningfulness of the general implication of the result.

In Ref. \cite{hp} we have shown that the scaling behavior of $P(S)$ is
insensitive
to the lattice size $L$, the nucleation radius $S_0$, and the nucleation
probability
$p$. Here, in this paper we have demonstrated its insensitivity to the
drift speed and
fluctuations, and have shown that the primary mechanism responsible for
scaling is
coalescence, not growth.  By adopting a set of simpler rules, we have
further
shown the independence on the factors (such as surface tension) that govern
the
shape of the clusters.  Moreover, the scaling exponent is independent of
whether
the clusters break up when there is sufficient nucleation.  The
universality of the
scaling behavior is truly amazing.

The independence on the various factors examined suggests that so long as
the
region in which hadrons are formed while in the environment of quarks and
gluons is large enough, universal scaling will ensue whatever the dynamical
details may be.  This is of tremendous significance to the search of
signatures of
phase transition when the confinement process is not well understood.  Of
course,
the realistic experimental signature still awaits further study on how the
hadronic
clusters produced at high temperature (considered here) are transformed to
cold hadrons at the detector.  Our result provides the motivation to
examine that
problem both theoretically and experimentally.  The result also suggests
the
possibility that the problem of self-organized criticality studied here may
in some
way be related to percolation in nuclear fragmentation.  To set the stage
for
connecting quark-hadron phase transition in heavy-ion collision with
self-organized criticality is in itself perhaps the most interesting
development that
this line of investigation has initiated.

\begin{center}
\subsection*{Acknowledgment}
\end{center}

This work was supported in part by the U.S. Department of Energy under
Grant No. DE-FG06-91ER40637, and by the Natural Science and Engineering
Council of Canada and by the FCAR fund of the Quebec Government.

\newpage
\begin{center}
\subsection*{Figure Captions}
\end{center}

\begin{description}
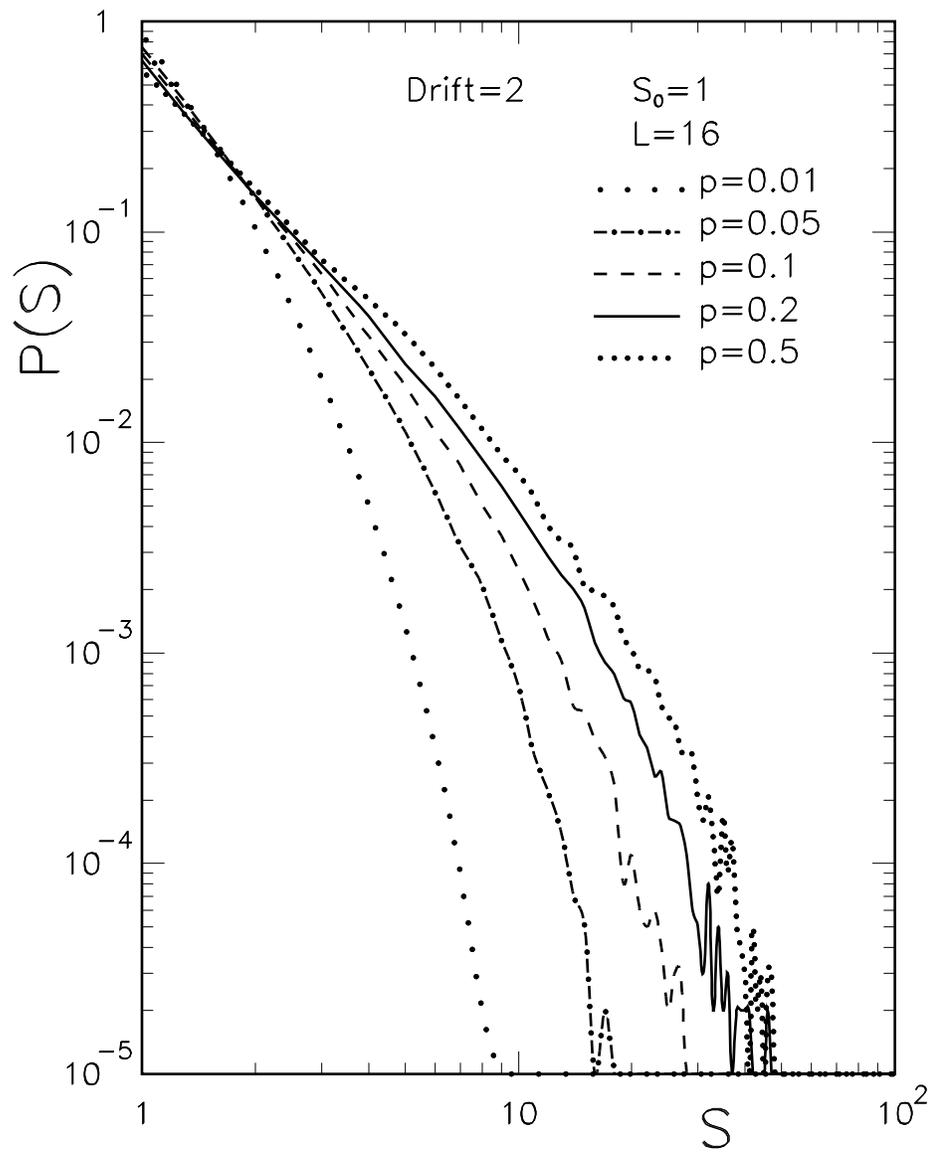
\item[Fig. 1]Cluster-size distribution $P(S)$ for $L = 16$ when the drift
step-size
is doubled.
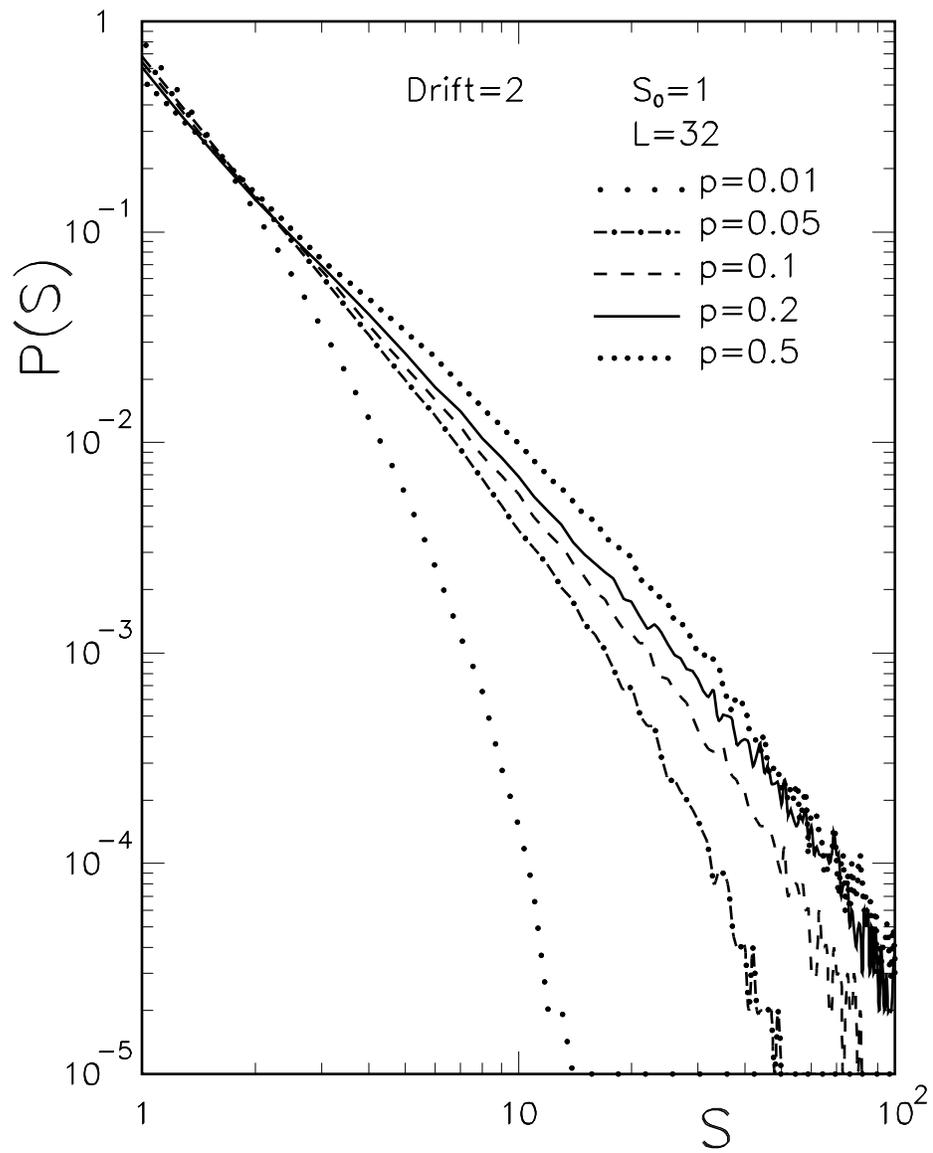
\item[Fig. 2]Same as Fig. 1, but for $L = 32$.
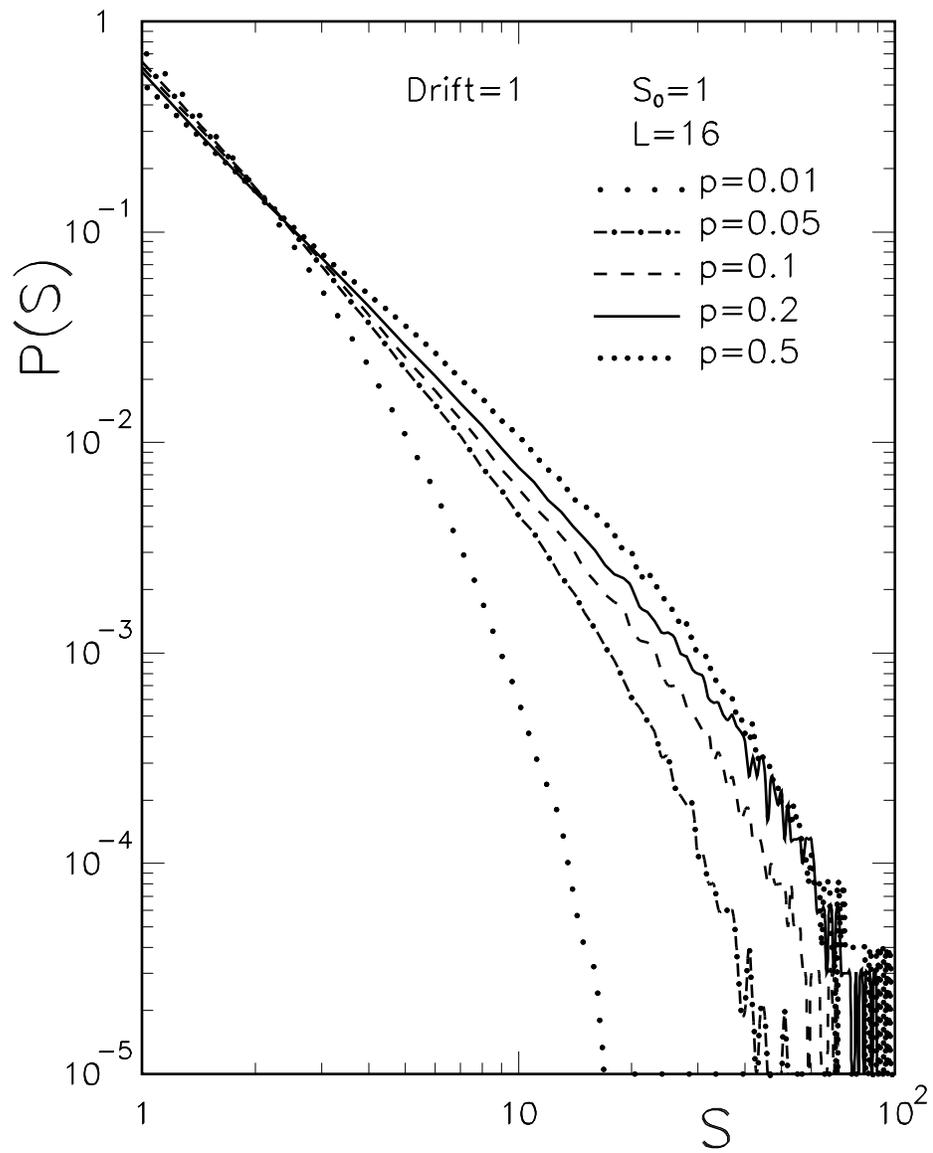
\item[Fig. 3]$P(S)$ for the standard case of $S_0 = 1$, $L = 16$ when the
drift
step-size is 1 unit.
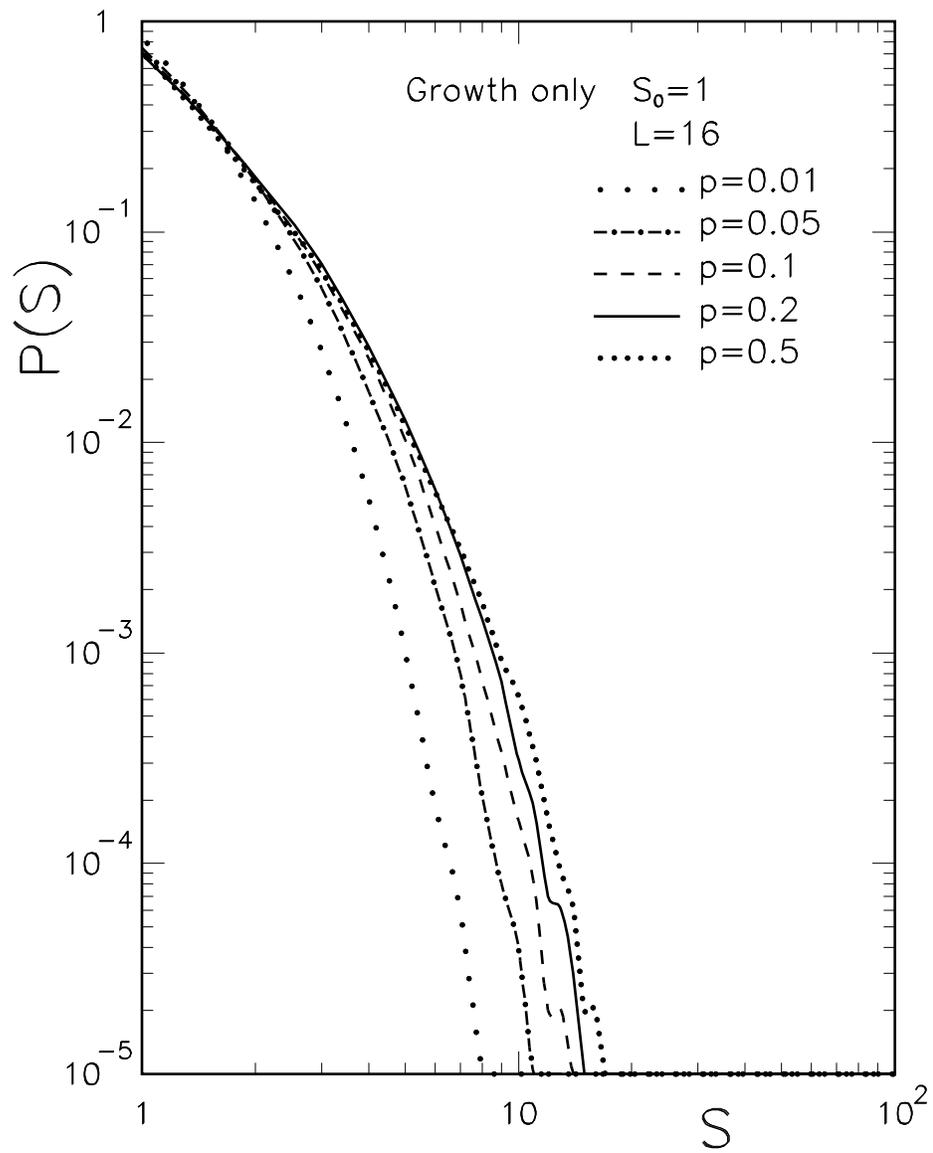
\item[Fig. 4]The standard case with growth only, without coalescence.
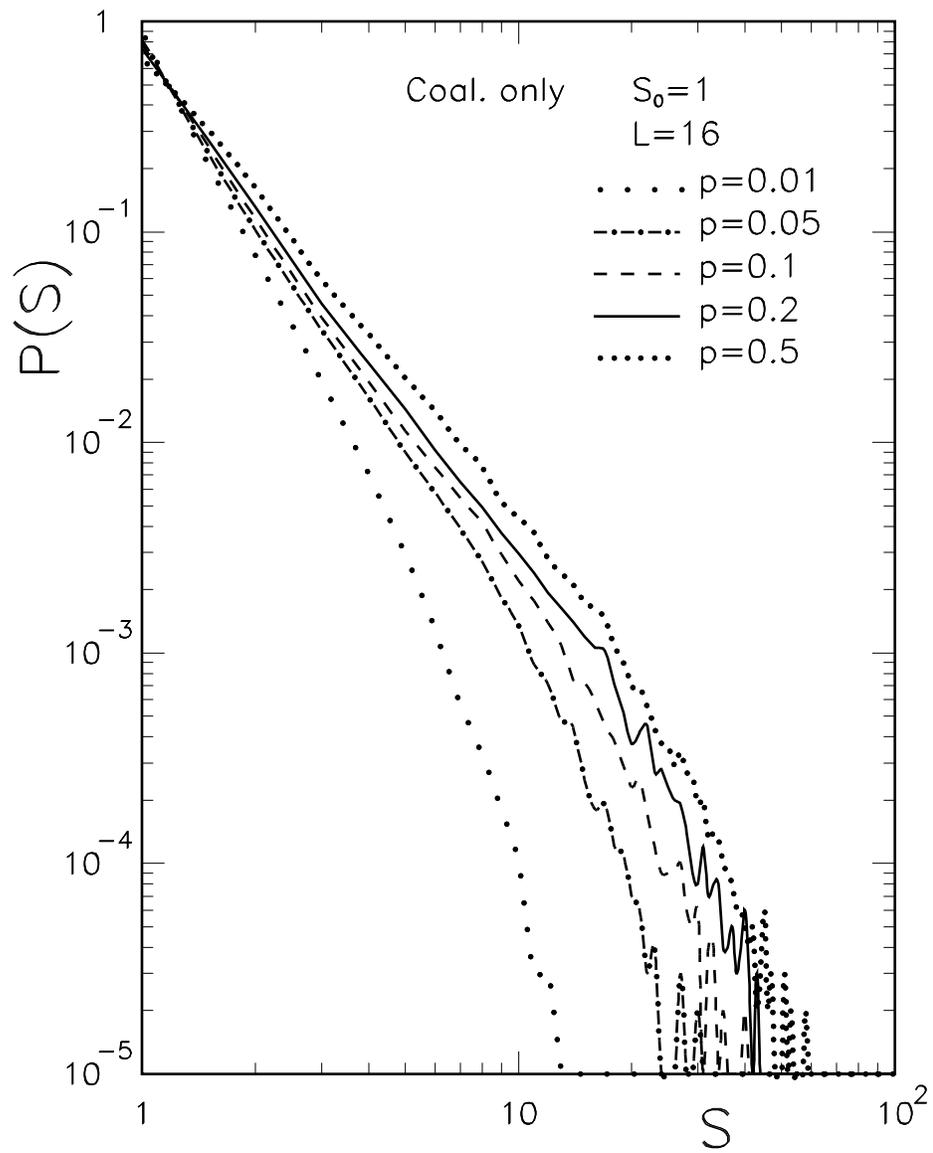
\item[Fig. 5]The standard case with coalescence only, without growth.
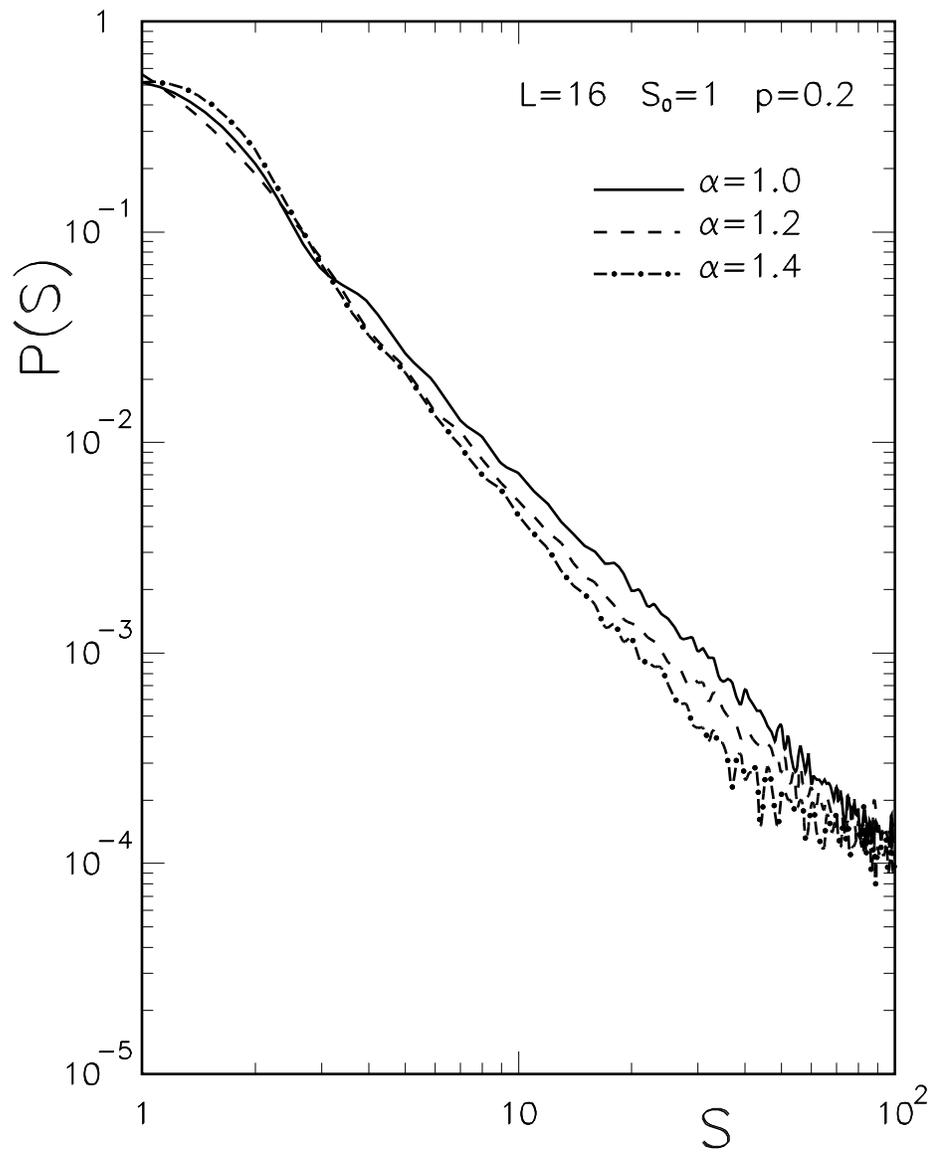
\item[Fig. 6]$P(S)$ for various effective radii parameterized by $\alpha$.
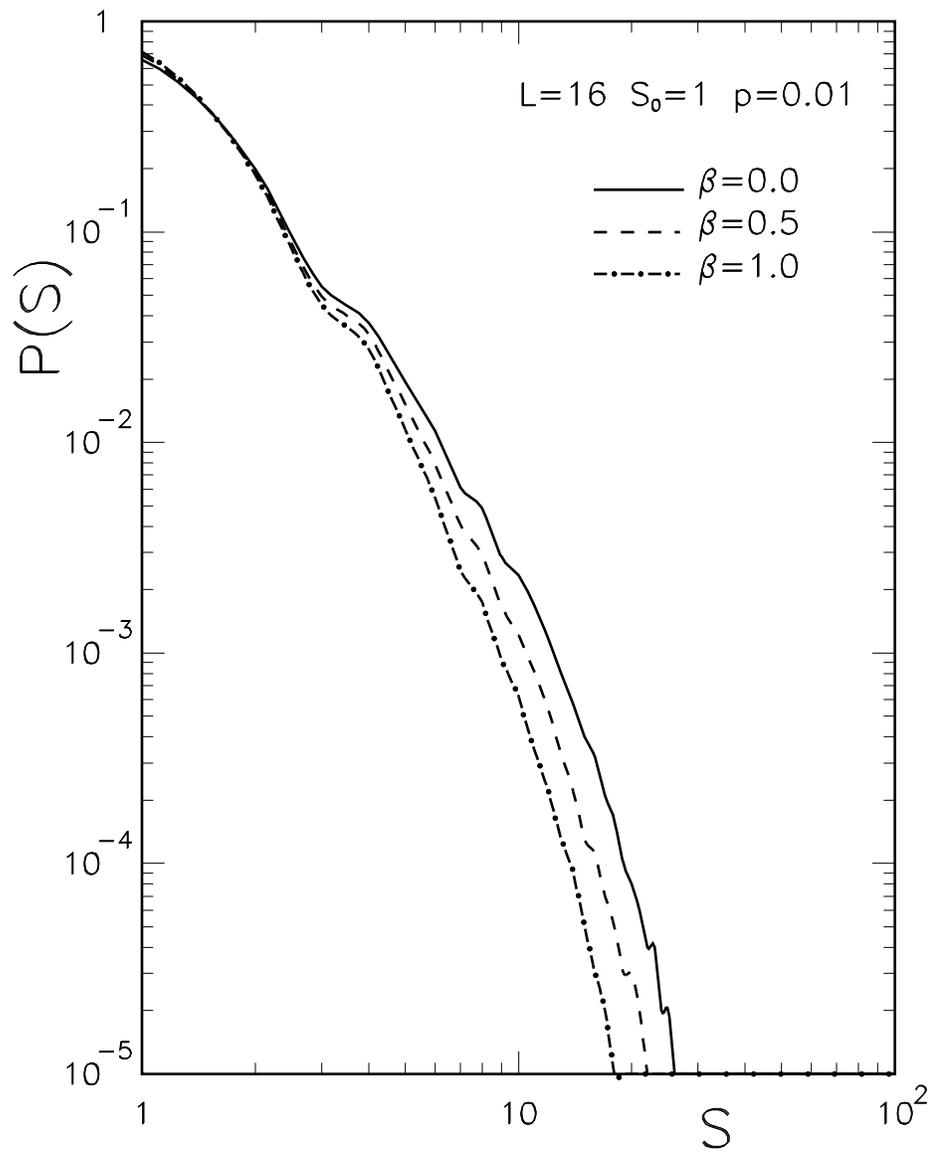
\item[Fig. 7]$P(S)$ for $p = 0.01$ when there is breakup parameterized by
$\beta$.
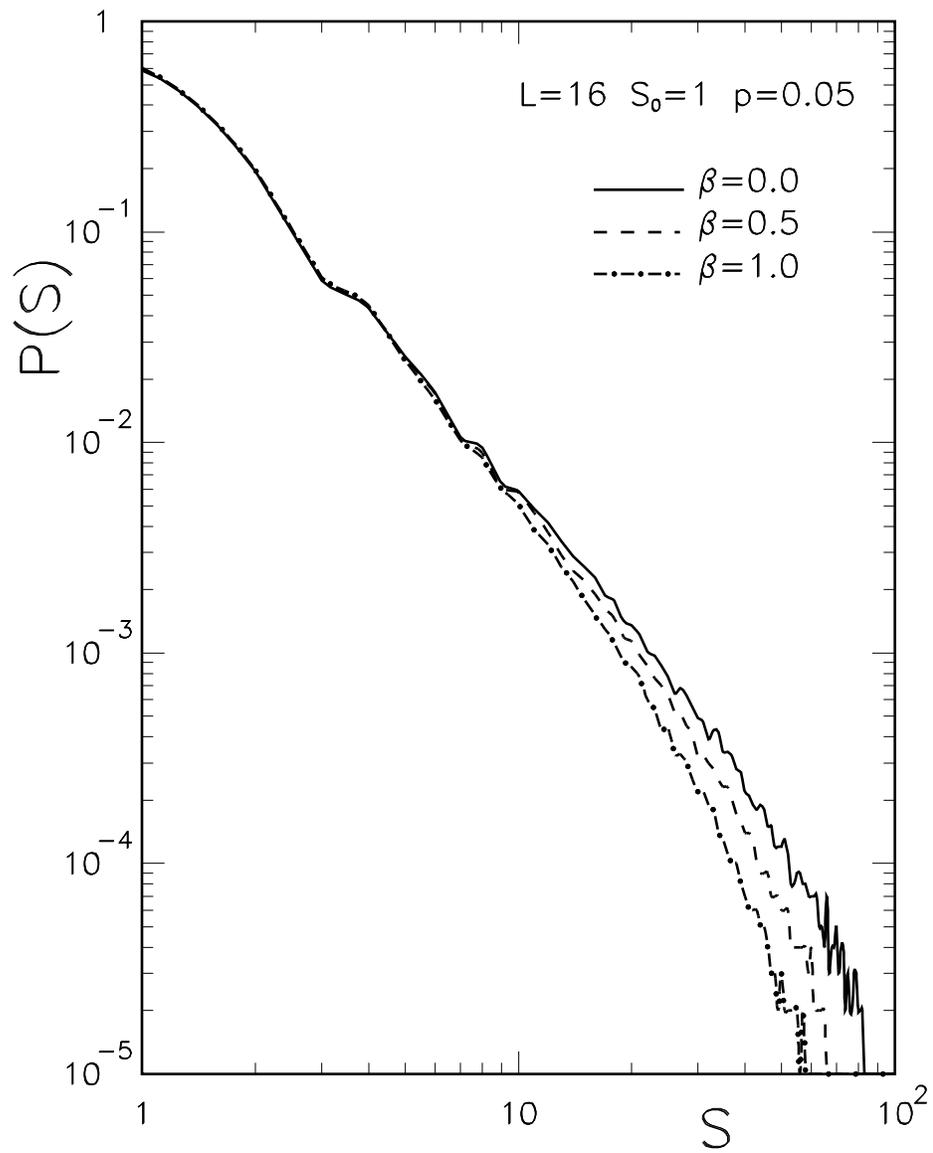
\item[Fig. 8]Same as Fig. 7, but for $p = 0.05$.
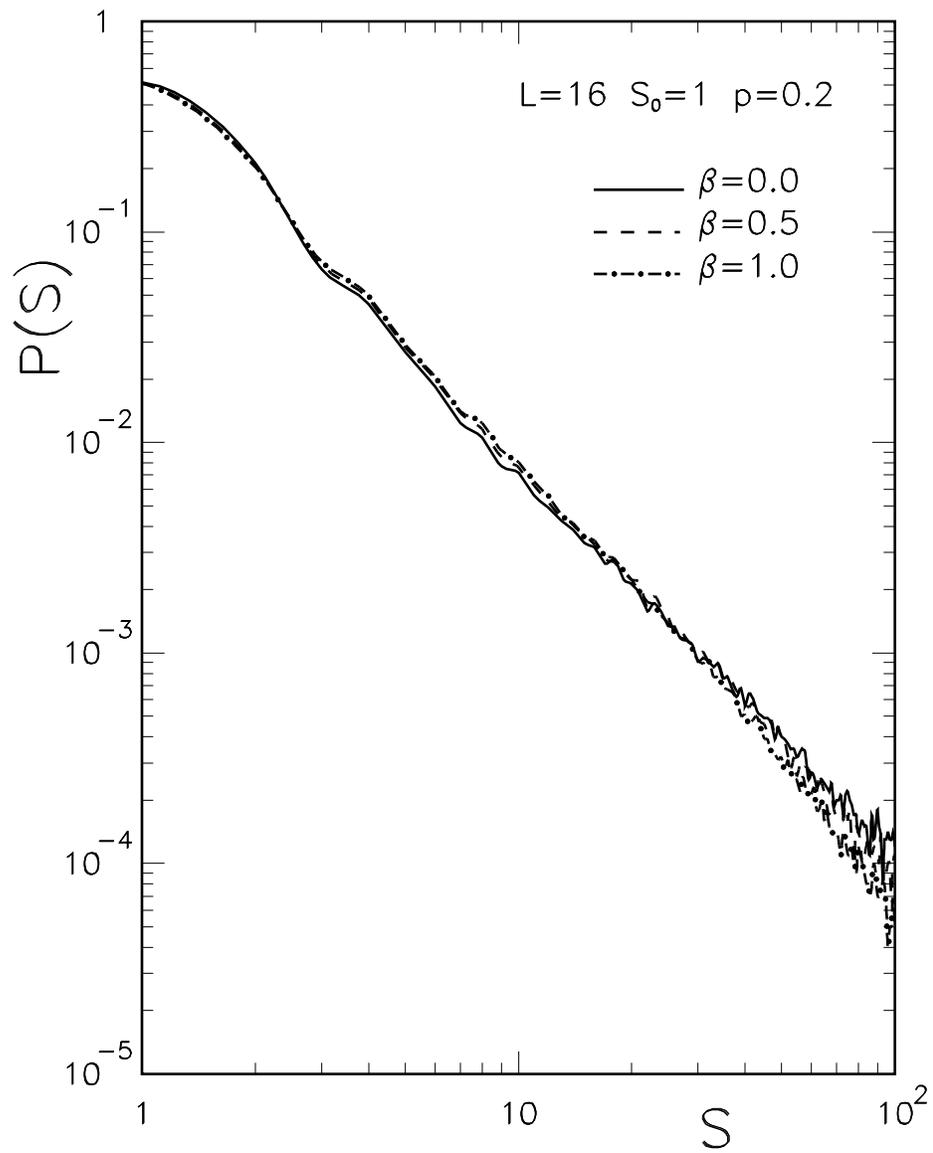
\item[Fig. 9]Same as Fig. 7, but for $p = 0.2$.

\end{description}

\end{document}